\documentclass[prl,twocolumn,amsmath,amssymb]{revtex4-1}% APS journal style
\usepackage{amsmath,amssymb,amsfonts,bm}
\usepackage{graphicx}
\usepackage{color}
\usepackage{bbold}
\usepackage{graphicx}% Include figure files
\usepackage{dcolumn}% Align table columns on decimal point
\usepackage{epstopdf}
\usepackage{color}
\usepackage{epsfig}
\usepackage{url}
\usepackage{hyperref}
\usepackage{verbatim}
\usepackage{enumerate}
\newcommand{\Tr}{\operatorname{Tr}}

\newcommand{\be}{\begin{equation}}
\newcommand{\ee}{\end{equation}}
\newcommand{\ba}{\begin{aligned}}
\newcommand{\ea}{\end{aligned}}

\newcommand{\john}[1]{{\color{green} #1}}

\usepackage{braket}

\newcommand{\aveeq}[1]{\langle #1 \rangle}

\newcommand{\titleinfo}{Eigenstate Correlations, Thermalization and the Butterfly Effect
}

\graphicspath{ {./figs/} }

\begin{document}

\title{\titleinfo}
\author{Amos Chan,  Andrea De Luca and J. T. Chalker}
\affiliation{Theoretical Physics, Oxford University, Parks Road, Oxford OX1 3PU, United Kingdom}

\date{\today}

\begin{abstract}
We discuss eigenstate correlations for ergodic, spatially extended many-body quantum systems, in terms of the statistical properties of matrix elements of local observables. While the eigenstate thermalization hypothesis (ETH) is known to give an excellent description of these quantities, the phenomenon of scrambling and the butterfly effect imply structure beyond ETH. We determine the universal form of this structure at long distances and small eigenvalue separations for Floquet systems. We use numerical studies of a Floquet quantum circuit to illustrate both the accuracy of ETH and the existence of our predicted additional correlations. 
\end{abstract}

\maketitle

\paragraph{Introduction.} Statistical mechanics is one of the pillars of modern physics.
There is however a well-known disjuncture between the fundamental laws of classical and quantum mechanics, 
%which predict a 
with reversible time evolution, and the description 
%possibility of describing macroscopic features 
of generic systems using 
%a small set of 
%equilibrium 
ensembles 
%in which a
%A sufficient condition for statistical equilibrium in an isolated system is that the 
%probability distribution is a function 
defined only by a small number of conserved quantities. %, such as total energy and total particle number.
Within classical mechanics, a standard supporting argument 
%in favour of this postulate 
is the ergodic hypothesis. A chaotic system 
%is one that 
evolves over time to explore uniformly all states
%``all accessible'' states: all those 
compatible with the conservation laws, so that 
%Hence, in an ergodic system, 
the microcanonical ensemble is the only possible equilibrium ensemble with fixed energy. 
%This provides an
%On the other hand, for an integrable system, the phase space is foliated by invariant tori almost everywhere, and we do not expect it to thermalize. These results provide a satisfying explanation of statistical behaviour in classical systems exhibiting chaos.
%
For quantum systems, the 
%formulation of a satisfying 
notion of ergodicity is 
%much 
more problematic and even the definitions of quantum chaos and integrability are under debate~\cite{Haake, caux2011remarks}. 

A fruitful direction is to characterise quantum systems in terms of 
%Progress has come from formulating questions about ergodicity in terms of
%An important step forward has been achieved when the questions about ergodicity have been posed in the corresponding language of 
the spectral properties of their Hamiltonians or evolution operators. %, generated either by an Hamiltonian or by its generalizations in Floquet systems. 
Here, 
%this approach has clarified the role of 
random matrix theory (RMT) 
%in providing 
provides an important paradigm \cite{Mehta}. Quantum chaotic systems can be identified, following the
% a successful application of this framework is the celebrated 
 Bohigas-Giannoni-Schmidt conjecture~\cite{bohigas1984characterization},  from 
 %statistical properties of their 
 an RMT eigenvalue distribution \cite{guhr1998random}, while the Berry conjecture \cite{berry1977mv} proposes that their eigenfunctions can be understood 
 %in the semiclassical limit 
 as a random superposition of plane waves. %, with Gaussian random coefficients. % distributed as Gaussian random variables.
 %, which allows identifying quantum chaotic systems by looking at the level spacing distribution~\cite{guhr1998random}. 
 Building on these foundations, the eigenstate thermalization hypothesis (ETH) links the properties of eigenvectors for many-body systems to statistical mechanics and the dynamics of equilibration
 % has provided a direct connection between the statistical properties of the energy eigenvectors and the capability of the system to dynamically relax to an equilibrium ensemble
 ~\cite{Deutsch, Srednicki, rigol2008thermalization}. 
%The ETH originates from the \textit{Berry's conjecture}, a condition for Hamiltonian eigenfunctions for quantum particles systems having a chaotic classical limit~\cite{berry1977mv}. It essentially states that the energy eigenfunctions behave as a superposition of plane waves, with coefficients distributed as Gaussian random variables. 
%More recently, the ETH is formulated in more general terms, by stating that the expectation value of a local observable in an energy eigenstate is the same as the thermal value with the same macroscopic properties. To date, the ETH
It constitutes a widely accepted expression of the notion of ergodicity for many-body quantum systems~\cite{rigol2012alternatives, Rigol}.
% and its violations have been used to characterise different types of quantum ergodicity breaking~\cite{rigol2009breakdown, santos2010onset, biroli2010effect, pal2010many, alba2015eigenstate}. %RMT automatically verify the ETH and have been used as an idealised yardstick for non-integrable many-body lattice models on the energy scale of the level spacing. However, one of the main limitations of ETH is that it cannot capture any spatial structure present in the underlying system.

By design, the ETH omits any spatial structure present in the underlying system. Our aim in this paper is to understand universal features of eigenfunction correlations in ergodic many-body systems that follow from spatial structure and lie outside the ETH. 

A single-particle counterpart to the questions we address is provided by studies of eigenfunction correlations in the metallic phase \cite{Thouless74} or at the Anderson transition \cite{Gefen,ChalkerDaniel,Chalker1990,FyodorovMirlin}, in models of disordered conductors without interactions. In that case behaviour is controlled by conservation of probability density. By contrast, for interacting systems it is the dynamics of quantum information that determines long-distance correlations.

%Another characterisation of chaos has been recently formulated in direct connection with 
A direct characterisation of such dynamics is provided by
the butterfly effect~\cite{kitaev2015simple, Gu2017, MSS}. In the framework of quantum mechanics, this phenomenon concerns the influence of a perturbation induced by the operator $\hat Y$ on later measurements of $\hat X$. This is quantified by the value of the commutator $[\hat X(t), \hat Y (0)]$, and an indication of the strength of the effect is given by
\begin{equation}
C(t) = \frac{1}{2}\aveeq{[ \hat X(t) , \hat Y(0) ]^\dagger[ \hat X(t) , \hat Y(0) ]}
\end{equation}
where $\aveeq{\ldots}$ denotes the thermal average. If the perburbation $Y$ and the measurement $X$ occur at points which are separated in space, the commutator is initially vanishing. More precisely, for short-range interactions in spatially extended lattice models, the Lieb-Robinson bound~\cite{Lieb} ensures that $C(t)$ remains exponentially small for a time which grows linearly in the space separation $\ell$ between the supports of $Y$ and $X$.
In this language, the phenomenon of scrambling is that $Y$ necessarily influences $X(t)$ at large $t$, and $C(t)$ approaches
$ \aveeq{\hat Y \hat Y}\aveeq{\hat X \hat X}$ regardless of the specific choice of $\hat Y$ and $\hat X$. 

An important recent insight \cite{FoiniKurchan2018} is that non-zero $C(t)$ implies correlations in matrix elements of operators beyond those captured by the ETH. We show here that these correlations acquire a specific universal form for pairs of widely-separated local operators in spatially extended systems. Moreover, since the time-scale for propagation of quantum information is long when $\ell$ is large, these additional correlations may be arbitrarily sharp in energy, in contrast to those of the ETH. 

We focus on Floquet systems because they constitute the simplest class. More generally, conservation laws are reflected in correlations at large distances and long times, and the simplest systems are ones with no conserved densities. To escape conservation of energy density, it is necessary to consider evolution with a time-dependent Hamiltonian, and for there to be a fixed evolution operator this time dependence should be periodic.

A convenient way to construct models with time-dependent evolution operators is by using unitary quantum circuits. These have yielded valuable insights into chaotic quantum dynamics both for systems with an evolution operator that is stochastic in time~\cite{Nahum2017, Nahum2017a, vonKeyserlingk2017} and for Floquet systems \cite{chan2017solution, chan2018spectral, amosmeasure}. In particular, these studies have confirmed the existence of two regimes for $C(t)$, distinguished by the sign of  $v_B |t| - \ell$, where the \textit{butterfly velocity} $v_B$ characterises the speed at which operators spread in space. We use Floquet random unitary circuits in the following for numerical simulations.

%This kind of manifestation of quantum chaos are generic and results from a combination
%of spatial structure and locality: they are therefore a perfect candidate to explore generic features of chaos which go beyond ETH. 
%
%Our aim in this letter is to identify the generic structure of correlations between eigenfunctions of chaotic many-body systems with interactions that are local in space.

\paragraph{ETH and relaxation.} We start by recalling the formulation of the ETH and its connection to the autocorrelation function of an observable. 
%Consider a chaotic many-body quantum system evolving in time with local interactions. For definiteness, we focus on one dimensional Floquet system and we denote with $\hat W$ 
%the unitary operator which evolves the state of one period. 
%Our considerations apply for higher dimensional systems and fixed Hamiltonians as well (see below). 
Consider a chaotic many-body Floquet system with local interactions and Hilbert space dimension $N$. Let  $\hat{W}$ be the evolution operator for one period, with 
%We indicate with greek letters the 
eigenstates $\ket{\alpha}$ 
%of $\hat W$ 
and eigenphases $E_\alpha$.
% the corresponding eigenphases.
According to the ETH, the matrix elements of a local Hermitian operator $\hat X$ have the form~\cite{srednicki1999approach}
\begin{equation}
\label{ethmatrix}
X_{\alpha\beta} \equiv \bra{\alpha} \hat X \ket{\beta} = \bar X \delta_{\alpha\beta} + {N}^{-1/2} h(\Delta_{\alpha\beta}) R_{\alpha\beta}^{X}
\end{equation}
where $\Delta_{\alpha\beta} = E_\alpha - E_\beta$ modulo $2\pi$.
In this expression, $\bar X$ is 
%the typical value of diagonal matrix elements and is 
the value to which the expectation of $\hat X$ relaxes at long times~\cite{biroli2010effect,brandino2011quench,kim2014testing}, $h(\omega)$ is a smooth function of  $\omega$, and $R_{\alpha \beta}$ are Gaussian random variables with zero mean and unit variance, which are complex and independent for each pair $\alpha > \beta$, and real and independent for each $\alpha=\beta$. Hermiticity of $\hat X$ implies that $h(\omega)$ is real and symmetric, and that $R_{\alpha \beta}^\ast = R_{\beta\alpha}$. Without loss of generality, we consider traceless operators so that $\bar X = 0$. It is then useful to define
\begin{equation}\label{F}
F(\omega) 
%\equiv (2\pi)^{-1}{[h(\omega)]^2}
= {N}^{-1} \bigg[\sum_{\alpha\beta} |X_{\alpha\beta}|^2 \delta(\Delta_{\alpha\beta} - \omega)\bigg]_{\rm av}
\end{equation}
with the eigenstate average $[ \ldots ]_{\rm av}$ either effected by broadening the delta function, or taken over an ensemble of statistically similar systems.
Using \eqref{ethmatrix} and the mean level spacing $\Delta= 2 \pi / N$, we write $F(\omega)\equiv (2\pi)^{-1}{[h(\omega)]^2_{\rm av}}$. $F(\omega)$ characterizes relaxation of the (integer $t$) autocorrelation function\footnote{The $\delta$-function is smeared on the scale of level spacing, which is exponentially small in system size. So Eq.~\eqref{autocorr} is exact in the thermodynamic limit.}, since
\begin{equation}
\label{autocorr}
\aveeq{X(t) X} = \int_0^{2\pi} \!\!\! {\rm d}\omega \, e^{ \imath \omega t} F(\omega) \equiv \aveeq{X^2} f(t) \,,
\end{equation}
where the thermal average appropriate for a chaotic Floquet system is the infinite-temperature one, with $\aveeq{ \ldots } \equiv {N}^{-1} \Tr[\,\ldots\,]$. 
Decay of the autocorrelation function on a microscopic relaxation timescale $t_R$ is encoded in $f(t)$, which satisfies $f(0)=1$ and $f(t) \to 0$ for $t \gg t_R$. 
%The scale $t_R$ is a microscopic quantity, expected to be independent of system size.
%This implies
%\begin{equation}
%F(\omega) \sim \left\{ \begin{array}{cl}  \aveeq{X^2} \, t_R & \quad |\omega t_R| \ll 1 \\ &\\0 & \quad |\omega t_R| \gg 1\,.
%\end{array} \right.
%\end{equation}

\paragraph{Generic form of four-point correlators.}
Let $\hat X$ and $\hat Y$ be local observables acting near the points $x$ and $y$, with $\ell \equiv |x - y|$. % \gg v_B t_R$. 
In analogy with Eq.~(\ref{F}), we introduce the correlator of four matrix elements
\begin{multline}\label{G}
G(\omega_1,\omega_2, \omega_3) = {N}^{-1} \bigg[\sum_{\alpha\beta\gamma\delta} X_{\alpha \beta} Y_{\beta \gamma} X_{\gamma \delta} Y_{\delta \alpha} \,\\
\delta(\Delta_{\alpha\beta} - \omega_3)\delta(\Delta_{\beta\gamma} - \omega_2)\delta(\Delta_{\gamma\delta} - \omega_1)\bigg]_{\rm av}\,.
\end{multline}

If one assumes [following Eq.~\eqref{ethmatrix}] that the matrix elements of $\hat X$ and $\hat Y$ are uncorrelated random variables, then 
 individual terms $X_{\alpha\beta}Y_{\beta\gamma}X_{\gamma\delta}Y_{\delta\alpha}$ are ${\cal O}(N^{-2})$ with random phase, and
only the term $\alpha=\beta=\gamma=\delta$ survives the average in Eq.~(\ref{G}), implying
% and we would have 
$G(\omega_1,\omega_2, \omega_3)\sim {\cal O}(N^{-2})$. 
%Moreover, before averaging we would expect the sum of $N^4$ terms \andrea{with random phase} to yield an ${\cal O}(N^{-1})$ contribution to $G(\omega_1,\omega_2, \omega_3)$. 
In fact, a sum rule shows that $X_{\alpha \beta} Y_{\beta \gamma} X_{\gamma \delta} Y_{\delta \alpha}$ has a coherent ${\cal O}(N^{-3})$ component, as well as ${\cal O}(N^{-2})$ fluctuations~\footnote{See supplementary material at [url].} (see also Ref.~\cite{FoiniKurchan2018}). We will show that for large $\ell$ this coherent component is concentrated on the terms in which the two matrix elements of $\hat X$ (and also those of $\hat Y$) are between pairs of states with almost opposite eigenphase differences. Specifically, the condition for  $X_{\alpha \beta} Y_{\beta \gamma} X_{\gamma \delta} Y_{\delta \alpha}$ to make a large contribution to $G(\omega_1,\omega_2, \omega_3)$ is that $(E_\alpha - E_\beta) \approx  -(E_\gamma - E_\delta)$, implying also $(E_\beta - E_\gamma) \approx -(E_\delta - E_\alpha)$.

To understand the form of $G(\omega_1,\omega_2, \omega_3)$ we switch to the time domain and consider $C_\ell(t) \equiv \frac{1}{2}\langle | [{\hat X}(t),{\hat Y}] |^2 \rangle$, written as
\begin{align}
\label{corr2}
C_\ell(t)  =   \langle \hat Y^2 \hat X^2(t)\rangle  - \langle \hat X(t) \hat Y \hat X(t) \hat Y \rangle \,.
%\\ &= 2 \langle X^2 \rangle \langle Y^2 \rangle g(|t| - |x-y|/v)\,.
\end{align}
The second term on the right-hand side is the \textit{out-of-time order correlator} (OTOC) \cite{LarkinOvchinnikov, MSS}.
$C_{\ell}(t)$ vanishes for short times $t$ and large separations $\ell$, while for
%as the operators $\hat X$ and $\hat Y$ commute. For 
large times the OTOC is small and $C_{\ell}(t)$ approaches $\aveeq{\hat X^2}\aveeq{\hat Y^2}$. Correspondingly,  the OTOC has the form
$\langle \hat X(t) \hat Y \hat X(t) \hat Y \rangle = \aveeq{\hat X^2}\aveeq{\hat Y^2} k_\ell(\ell/v_B - |t|)$, where $k_\ell(\tau)$ steps between $k_\ell(\tau) =1$ for $\tau$ large and positive, and $k_\ell(\tau)=0$ for $\tau$ large and negative. The width $\Delta \tau$ of the step satisfies $\Delta \tau \ll v_B \ell$ for large $\ell$  \cite{Nahum2017a, vonKeyserlingk2017}.
% \john{the step is $\Delta\tau \sim \ell^{1/2}/D$  in one-dimensional systems, set}  by diffusive processes \cite{Nahum2017a, vonKeyserlingk2017}. 
%in analogy to what was done in \eqref{autocorr} for the autocorrelation, we encode this behaviour by setting
%\begin{equation}
%\label{OTOCg}
%C_\ell(t) = 2 \langle X^2 \rangle \langle Y^2 \rangle \left(1 - g\Bigl(\frac{\ell - v_B |t|}{v_B}\Bigr)\right)\,.
%\end{equation}
%Here, $v_B$ is the butterfly velocity and $g(\tau)$ is a function that varies between  $g(\tau)=0$ and $g(\tau) = 1$ as $\tau$ passes from $-\infty$ to $\infty$, with a step width much smaller than $t$. 

We want to connect this behaviour with the 
statistical properties of matrix elements. 
%spectrum and eigenvectors of the Floquet operator $\hat W$. 
We start by introducing a generalised OTOC
\begin{equation}
\label{defsmallg}
\langle X(t_3) Y(t_2) X(t_1) Y \rangle \equiv \langle X^2 \rangle \langle Y^2 \rangle g(t_1, t_2, t_3)\,.
\end{equation}
This reduces to the standard OTOC for  $t_1= t_3 \equiv t$ and $t_2 = 0$. 
We argue that the relation \begin{equation}\label{factorise}
g(t_1,t_2,t_3) \approx f(t_3{-}t_1) f(t_2) k_\ell (\ell /v_B {-} |t_1|)
\end{equation} 
holds for  $\ell \gg v_B t_R$, where for simplicity we take the autocorrelation function $f(t)$ of $\hat X$ and $\hat Y$ to be the same.

A simple justification is as follows (for details, see \cite{Note2}).
We consider three regimes. In (i) $|t_2| \gg t_R$ and/or $|t_3-t_1| \gg t_R$. In (ii) and (iii)  $|t_2| \lesssim t_R$ and $|t_3-t_1| \lesssim t_R$. In addition, in (ii) $\ell -v_B|t_1-t_2| \gg \Delta \tau$, while in (iii) $\ell - v_B|t_1-t_2| \ll -\Delta \tau$. In (i) the left side of (\ref{factorise}) is zero because of scrambling \cite{Note2}, as is $f(t_3-t_1)f(t_2)$ on the right side. In (ii) $[Y(t_2),X(t_1)]$ is small in norm
%. Then, if $|t_1 - t_3| \lesssim O(t_R)$,
and we are dealing with averages taken approximately simultaneously at $x$ and $y$, which can be factorised. Hence $\langle X(t_3) Y(t_2) X(t_1) Y \rangle \approx  \langle X(t_3) X(t_1) \rangle \langle Y(t_2) Y \rangle $ and $g(t_1,t_2,t_3) \approx f(t_3{-}t_1)f(t_2)$; since $k_\ell (\ell /v_B {-} |t_1|)=1$ in (ii), Eq.~(\ref{factorise}) follows. In (iii) we can factorise $\langle X(t_3) Y(t_2) X(t_1) Y \rangle \simeq \langle X(t_3) \rangle \langle Y(t_2) X(t_1) Y \rangle = 0$, with corrections that vanish in the limit $|t_3 - t_1|/t_R\gg 1$. In this regime $k_\ell (\ell /v_B {-} |t_1|)=0$, so again Eq.~(\ref{factorise}) is satisfied.
%Consider first $\ell -v_B |t_1 - t_2| $ large and positive \andrea{so that $[Y(t_2),X(t_1)]$ is small in norm. Then, if $|t_1 - t_3| \lesssim O(t_R)$, we are dealing with averages taken approximately simultaneously at $x$ and $y$, which can be factorised.} As a consequence, in this regime $\langle X(t_3) Y(t_2) X(t_1) Y \rangle \approx  \langle X(t_3) X(t_1) \rangle \langle Y(t_2) Y \rangle $ and so $g(t_1,t_2,t_3) \approx f(t_3{-}t_1)f(t_2)$. Moreover, in this regime, if $f(t_2)\not=0$ then  $k_\ell (\ell /v_B {-} |t_1|)=1$, yielding (\ref{factorise}). \andrea{If instead $t_3$ is far from $t_1$ so that $\ell -v_B |t_3 - t_2| < 0$, then we can factorise $\langle X(t_3) Y(t_2) X(t_1) Y \rangle \simeq \langle X(t_3) \rangle \langle Y(t_2) X(t_1) Y \rangle = 0$ up to exponentially small corrections in $|t_3 - t_1|/t_R$.} In the opposite limit,  with $\ell -v_B |t_1 - t_2|$ large and negative, both sides of Eq.~\eqref{factorise} are small.}

Corrections to Eq.~\eqref{factorise} are expected to be parametrically small except near the butterfly front: the regime excluded from (i) - (iii), in which  $|t_2| \lesssim t_R$, $|t_3-t_1| \lesssim t_R$ and $\left| \ell /v_B {-} |t_1| \right|\lesssim \Delta \tau$. This has a width in $t_1$ that is much narrower at large $\ell$ than the main scale $\ell/v_B$. 
%In particular, Eq.~\eqref{factorise} becomes an exact statement in the limit of large $\ell$ with $t_1 = \ell/v_B$,  $t_3 = t_1 + \tau$, $t_2 = \tau'$ and fixed $\tau, \tau'$.

Clearly $G(\omega_1,\omega_2, \omega_3)$ and $g(t_1, t_2, t_3)$ and are related by Fourier transform, obtained by summing over the times $t_1, t_2, t_3$. The corrections to \eqref{factorise} affect only a vanishing fraction of contributions for $\ell \gg v_B t_R$. From this, we deduce
\begin{align}
\label{Gform}
G(\omega_1,\omega_2, \omega_3) \overset{\underset{\ell \to \infty}{\lim}}{=}
F(\omega_2)
F(\omega_3)
K_\ell(\omega_1+\omega_3)\,,
\end{align}
where we have introduced the Fourier transform
\begin{equation}
K_\ell(\omega) = \frac{1}{2\pi} \sum_t k_\ell(\ell/v_B -|t|) e^{-i\omega t}\,.
\end{equation}
%Inspecting the expression in \eqref{Gform}, we see that it 
Our discussion of the form of $k_\ell(t)$ implies that $K_\ell(\omega)$ is maximum at $\omega= 0$ and has a width in frequency of order $v_B/\ell$. 
%For large  $\ell$, this is much narrower than the width $1/t_R$ of the function $F(\omega)$. 
At large $\ell$, since $\Delta \tau \ll \ell/v_B$, we can represent $k_\ell(\tau)$ as a step function and obtain the scaling 
form
\begin{equation}\label{k}
\lim_{\ell \to \infty} \frac 1 \ell K_{\ell}({u}/{ \ell}) = \frac{\sin (u/v_B) }{2\pi u}
\end{equation}
dependent only on the butterfly velocity of the model. 

Eqns.~(\ref{Gform}) and (\ref{k}) constitute our main theoretical results. They apply to a pair of operators acting at points separated by $\ell \gg v_B t_R$. In this limit they show that non-Gaussian correlations of matrix elements, which are not modelled by the ETH, have universal structure in frequency. This structure appears on a much finer scale ($v_b/\ell$) than the one ($1/t_R$) relevant for the Gaussian correlations that are represented by the ETH. 

%It is important to note (see also Ref.~\cite{FoiniKurchan2018}) that the scales of the Gaussian and non-Gaussian correlations are widely separated in ${N}$: individual terms $X_{\alpha \beta} Y_{\beta \gamma} X_{\gamma \delta} Y_{\delta \alpha}$ entering the sum in Eq.~(\ref{Gform}) are ${\cal O}({N}^{-2})$. If they had independent random phases, their sum would yield $G(\omega_1,\omega_2, \omega_3)\sim {N}^{-1}$. The finite value of $G(\omega_1,\omega_2, \omega_3)$ indicates that $X_{\alpha \beta} Y_{\beta \gamma} X_{\gamma \delta} Y_{\delta \alpha}$ has a coherent ${\cal O}({N}^{-3})$ component in addition to ${\cal O}({N}^{-2})$ fluctuations.

\paragraph{Model.} 
To test these ideas in a computational study, 
we consider a one-dimensional $L$-site Floquet unitary circuit~\cite{chan2017solution} generated by Haar-distributed random unitaries, where the quantum states at each site span a $q$-dimensional Hilbert space. The circuit is defined by a $q^L\times q^L$ Floquet operator $W=W_2\cdot W_1$, where $W_1 = U_{1,2} \otimes U_{3,4} \otimes  \ldots U_{L-1,L}$ and on an open chain $W_2 = \mathbf{1}_q \otimes U_{2,3} \otimes U_{4,5} \otimes  \ldots\mathbf{1}_q $. Here each $U_{i,i+1}$ is a $q^2 \times q^2$ random unitary matrix acting on sites $i$ and $i+1$. 
%For later convenience, 
We note that the circuit can be defined on a closed chain by the replacement $W_2 \to W_2 \otimes U_{1, L}$. 
This model has no conserved quantities or discrete symmetries. Moreover, many dynamical quantities can be computed analytically for $q \to \infty$~\cite{chan2017solution}: in this limit the autocorrelation function decays to zero in a single Floquet period ($t_R \to 0$) and the OTOC exhibits a light-cone with $v_B = 2$ and no broadening of the front. 
%In this limit, Eq.~\eqref{k} is accurate at finite $\ell$. 
At finite $q$, the Floquet random circuit provides an ideal setting to investigate the general phenomenology of Eqns.~(\ref{Gform}) and (\ref{k}).

\begin{figure}[t!]
	\includegraphics[width=0.95\columnwidth]{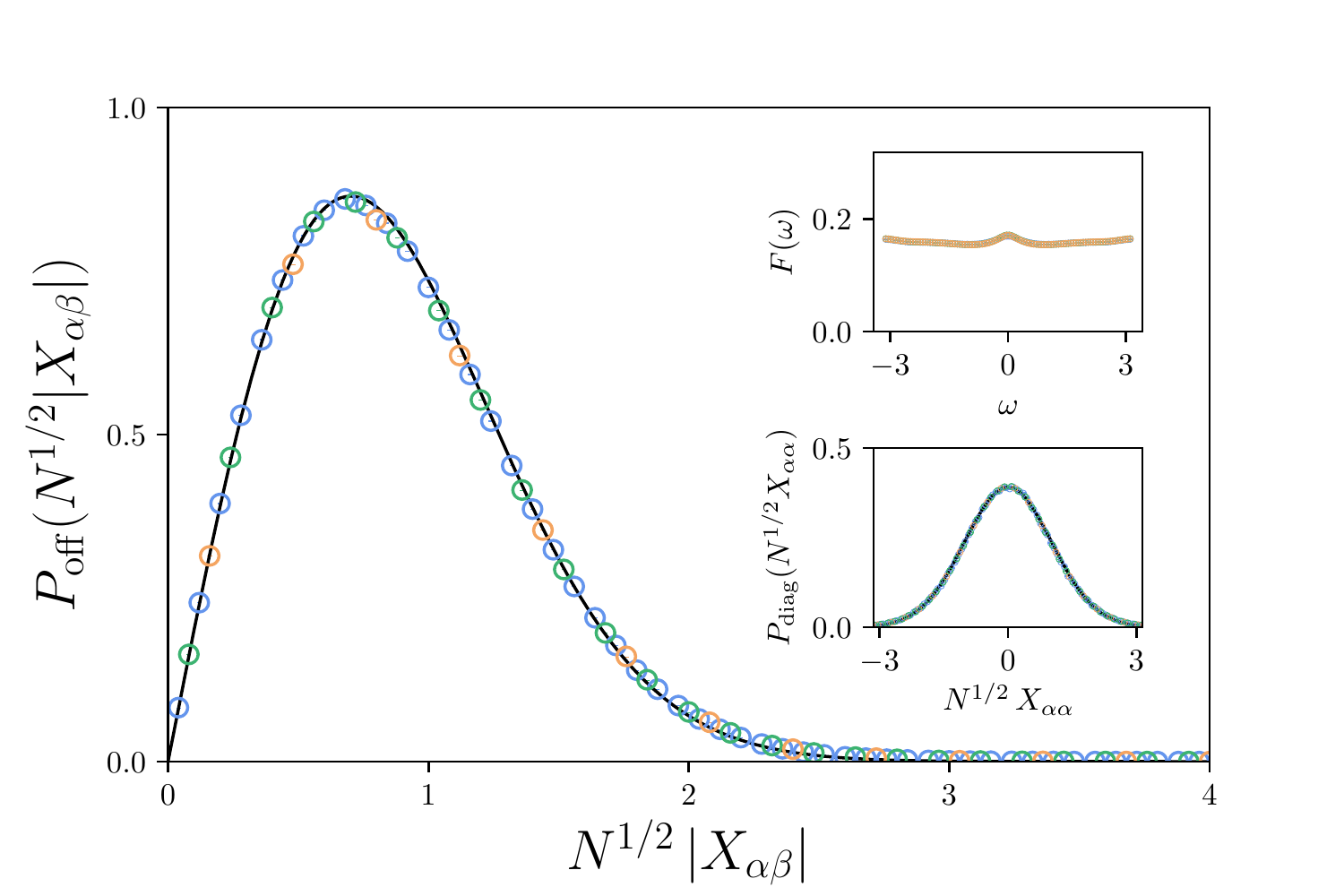}
	\caption{Comparison of predictions from the ETH with numerical results for Floquet quantum circuits (see text for definitions). 
	%The statistics of the 2-point correlator $X_{\alpha \beta}$ against the predictions of ETH (the ensemble average of $X_{\alpha \beta}$ is taken implicitly). 
	Main panel: Scale-collapsed probability distribution $ P_{\text{off}}(N^{1/2} |X_{\alpha \beta}|)$ of the modulus of off-diagonal elements $X_{\alpha,\beta}$ of a local operator,
		%vs $N^{1/2} |X_{\alpha \beta}|$ 
		%where $\alpha \neq \beta$, 
		with $N = 2^L$. Points: data for $L=6,8,10$ in blue, green, yellow respectively. Line: 
		%As expected from ETH, $X_{\alpha\beta}$ is a 
		complex Gaussian distribution, as expected from the ETH. % number whose modulus follows a Hoyt distribution, $h(x,\sigma ) \propto x \exp (- x^2 /2 \sigma^2)$. 
		Bottom inset:  Scale-collapsed probability distribution of the diagonal elements $P_{\text{diag}}(N^{1/2} X_{\alpha \alpha})$. Points: data for system sizes as in main panel. Line: real Gaussian distribution, as expected from ETH.
		Top inset: %The variance of the off diagonal elements 
		$F(\omega)$ vs $\omega$ %with $\alpha \neq \beta$ 
		for the same system sizes.  %Curves for different system sizes collapse under this scaling in our convention. 
		%Bottom inset:  The scale-collapsed probability distribution of the diagonal elements $P_{\text{diag}}(N^{1/2} X_{\alpha \alpha})$ for the same system sizes above. Again, as expected from ETH, $X_{\alpha \alpha}$ is a real Gaussian number.
	} \label{fig:2pt}
\end{figure}

\paragraph{Numerical simulations.}
We focus on $q=2$ where the space of single-site operators is spanned by Pauli operators $\hat \sigma_{\alpha}^j$, with $\alpha = x,y,z$ and $j = 1,\ldots, L$. We consider system sizes $L=4, 6, \dots, 12$ and perform the full diagonalization of the unitary matrix $\hat W$. 
We first study the statistics of the matrix elements $X_{\alpha \beta}$ and the function $F(\omega)$ for the operator $\hat X = \sigma_z^1$, using a closed chain in order to minimise boundary effects. The short-time behavior of the autocorrelation function can be computed analytically~\cite{chan2017solution} giving $\langle X(t)X\rangle = 1$, $0$, $6.67\times 10^{-3}$ and $3.87\times 10^{-3}$, for $t=0$, $1$, $2$ and $3$. Hence in this model $t_R$ is short and $F(\omega)$ is almost constant. For this reason, we neglect the dependence of $h(\omega)$ on $\omega$ and compute the combined probability distributions of all off-diagonal matrix elements, and of all diagonal elements.
As shown in Fig.~\ref{fig:2pt}, ETH gives an outstandingly accurate description for both quantities. (For similar results in a Hamiltonian system, see e.g. Ref~[\onlinecite{MoessnerHaque}, 
\onlinecite{luitz2016anomalous}]
)

\begin{figure}[t!]
	\includegraphics[width=0.85\columnwidth]{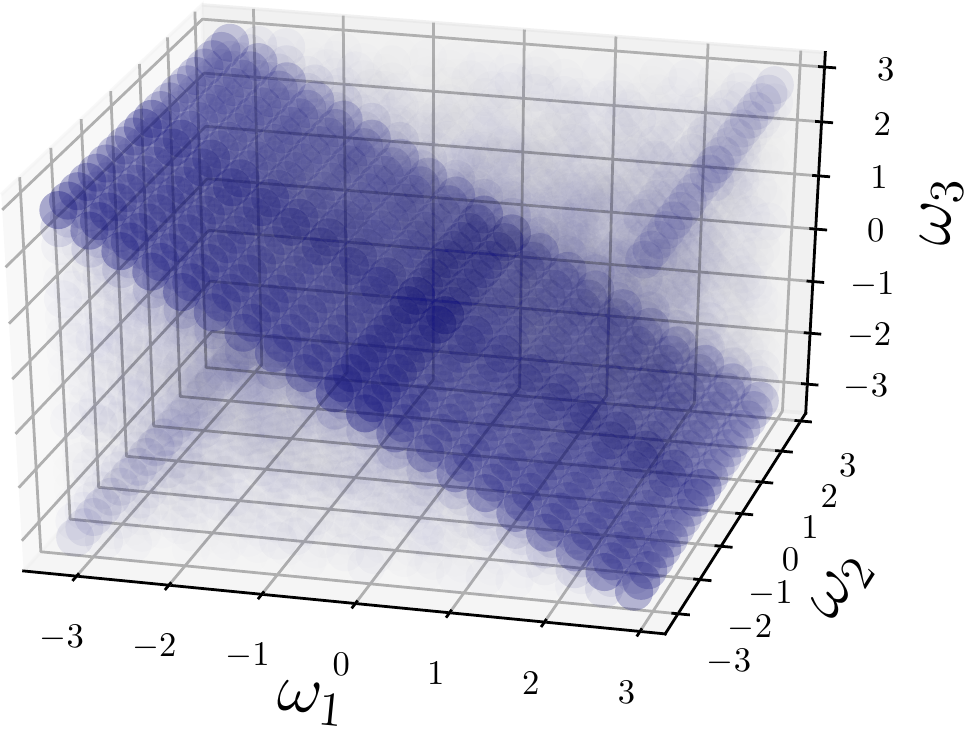}
	\caption{Histogram of $G(\omega_1, \omega_2, \omega_3)$ as a function of $\omega_1$, $\omega_2$ and $\omega_3$, for $L=12$ with 20 bins along each axis. %averaged \amos{around 700} %$10^3$ realisations. 
	Larger values of $|G(\omega_1, \omega_2, \omega_3)|$ are shown with heavier shading. 
	%There is a clear peak at $\omega_z =0$ along the $\omega_z$-axis and an   almost flat distribution along the $\omega_x$- and $\omega_y$-axis. The number of realizations is  $O(10^3)$. 
	} \label{fig:3d}
\end{figure}

Next we turn to the four-point correlators and test the predictions of Eqns.~(\ref{Gform}) and (\ref{k}). 
To maximise the separation $\ell = L-1$ at a given $L$,
we choose two operators acting on sites at opposite ends of an open chain:
\begin{equation}
\label{opdef}
\hat X = \hat \sigma_z^1\;,  \quad \hat Y = \hat \sigma_z^L \; .
\end{equation}
An overview of the data for $G(\omega_1,\omega_2,\omega_3)$ is given in Fig~\ref{fig:3d}. 
For the largest accessible system size, $L=12$, in each realization we sample $10^8$ contributions to each $\omega$ bin.
%$G(\omega_1,\omega_2,\omega_3)$ $10^8$ times.
 Additionally, we average $G(\omega_1,\omega_2,\omega_3)$ over around $700$ realizations.
The data show the expected narrow maximum near the plane $\omega_1+\omega_3=0$. For a quantitative analysis, we project $G(\omega_1,\omega_2,\omega_3)$ onto two orthogonal lines. First, we have as an identity (taking 
%the argument of the $\delta$-function 
$\omega_1+ \omega_3 -\omega$ mod $2\pi$)
\begin{equation}\label{I}
K_\ell(\omega) = \int_{[-\pi,\pi]^3}\!\!\!\!\!\!\!\!\!\!\!\!\!{\rm d}\omega_1 {\rm d}\omega_2 {\rm d}\omega_3 \, \delta(\omega_1+\omega_3-\omega) G(\omega_1,\omega_2,\omega_3)\,. \nonumber
\end{equation}
 Second, we define
\begin{equation}\label{J}
J(\omega_2) = \int_{[-\pi,\pi]^2}\!\!\!\!\! {\rm d}\omega_1 {\rm d}\omega_3 \, G(\omega_1,\omega_2,\omega_3)\,.
\end{equation}
From Eq.~(\ref{Gform}) we expect $J(\omega) \approx F(\omega)$. 
% we plot the histogram of $G(\omega_x, \omega_y, \omega_z)$ against $\omega_x$, $\omega_y$ and $\omega_z$ introduced in \eqref{coordchange}  for $L=12$ with 20 bins along each axis. The smaller the values of $G$, the less opaque the data are. It is clear how most of the weight is concentrated around $\omega_z \simeq 0$ with only a weak dependence on the other coordinates $\omega_x, \omega_y$. This is in consistent with the general behaviour predicted by \eqref{Gform} and the almost flat $F(\omega)$. 
%
%

\begin{figure}[t!]
\includegraphics[width=0.95\columnwidth]{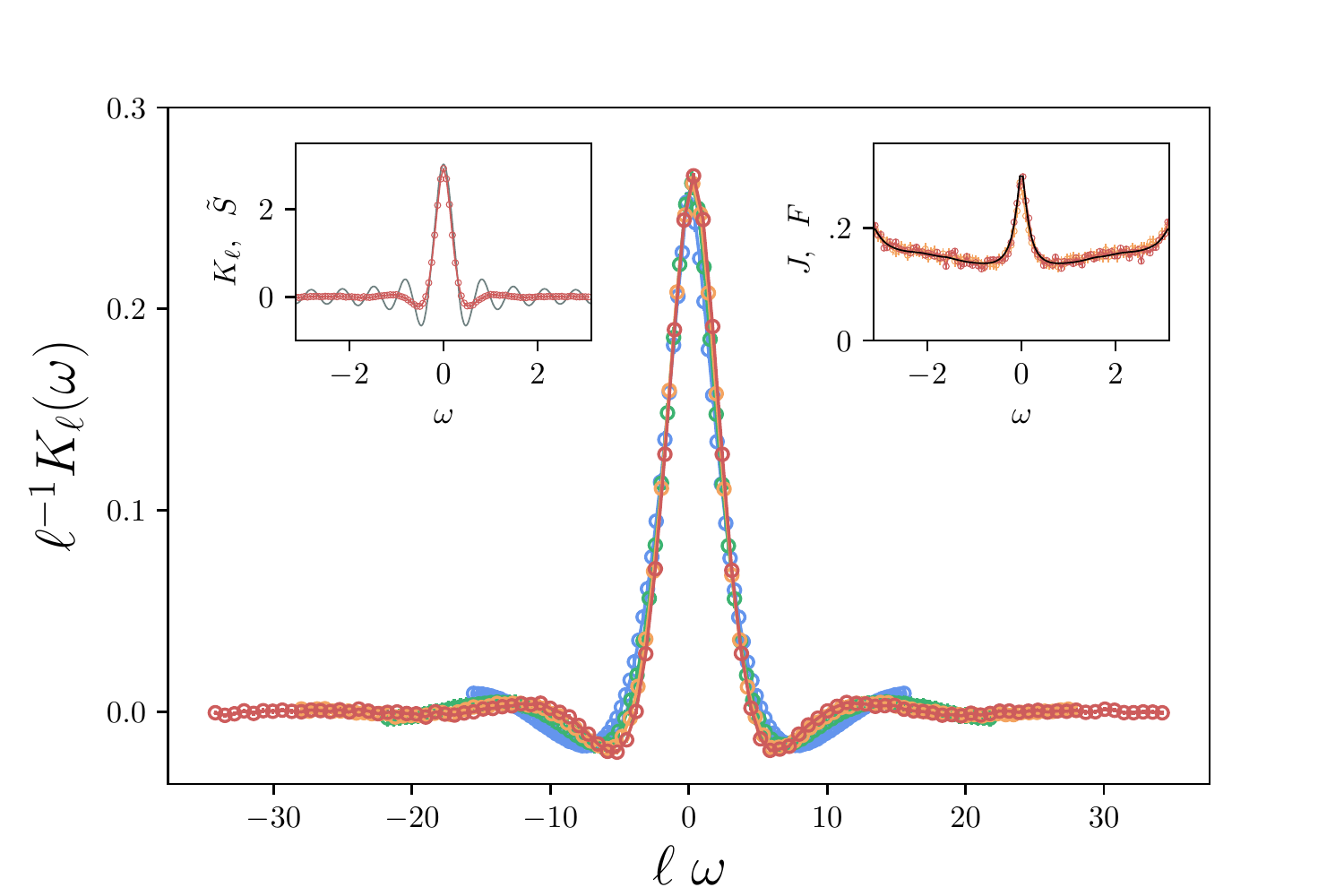} 
	\caption{%\textbf{[Caption to be updated]} Behaviour of $G(\omega_1,\omega_2,\omega_3)$ projected onto lines in $\omega$-space: the functions $I(\omega)$ and $J(\omega)$ [Eqns.~(\ref{I}) and (\ref{J})].
	Main panel: $\ell^{-1}K_\ell(\omega)$ vs $\ell \omega$ for $L=6,8,10,12$ in blue, green yellow and red respectively. 
	%$\ell^{-1} \mathcal{I}_{\omega_x, \omega_y}(\omega_z) := \ell^{-1} \int d \omega_x d \omega_y \;  G(\omega_x, \omega_y, \omega_z) $ vs. $\ell \omega_z$ for $L=6,8,10,12$. 
	Left inset: $K_\ell(\omega)$ vs $\omega$ (red) compared with Fourier transform $\tilde{S}$ of the step function $\Theta(|\ell/ v_b -t|)$ (grey) for $L=12$. 
	%See text for value of $v_b$. 
	Right inset:  
	$J(\omega)$ vs $\omega$ for  $L=10,12$ (yellow and red) compared with $F(\omega)$ vs $\omega$ for $\hat X$ acting on a site at the end of an open $L=12$ chain  (black). 
%	Right (from the top): To test Eq.~\ref{Gform}, we compare $G(\omega_1, \omega_2, \omega_3) $ vs $\omega_1$ (blue) with $F(\omega_2) F(\omega_3) K_\ell (\omega_1+ \omega_3)$ vs $\omega_1$ (red) for $L=12$ at $\omega_2 = -1.73$ and $\omega_3 = -1.41, -0.79, -0.16  $ respectively. 	Plots for other values of $\omega_2$ and $\omega_3$ show similar agreement.
	%$ \mathcal{I}_{\omega_x, \omega_z}(\omega_y) := \int d \omega_x d \omega_z \;  G(\omega_x, \omega_y, \omega_z) $ vs. $\omega_y$. The small peak scales trivially with $\ell$ and originates from the auto-correlation of local operators at the edge of a system with OBC (see its Fourier transform in the bottom right inset). 
	%Bottom right: $F(\omega)$ vs $\omega$ for $\hat X$ acting on site at end of open chain for $L=??$. 
	%The peak is more prominent than the one in the top right inset of Fig.~\ref{fig:2pt} with PBC.
		} \label{fig:4pt}
\end{figure}

Results for both functions are presented in Fig~\ref{fig:4pt}, and match excellently the expectations we have described.  
%we plot $ \ell^{-1} \mathcal{I}(\omega_z) :=  \int d\omega_x  \, d \omega_y  \, G(\omega_x, \omega_y, \omega_z) $ against $\omega_z / \ell$ for $L=6,8,10,12$.
The data in the main panel show a perfect collapse of the central peak for all accessible system sizes, in agreement with the scaling form of Eq.~\eqref{k}. 
The left inset shows a fit of the central peak in $K_\ell(\omega)$ to the Fourier transform of the step function $\Theta(|\ell/ v_B -t|)$. Since $t$ takes only integer values, the fit yields a range of possible values for the butterfly velocity: from $L=10$, we obtain $v_B\in [ 1.125, 1.286]$. This range includes the value for a random quantum circuit, $v_B = 2 (q^2 -1) / (q^2 +1)$ or $v_B=1.2$ at $q=2$~\cite{Nahum2017a,vonKeyserlingk2017}. The deviations of the data from the fitting function away from the central peak are due to the diffusive broadening of the step in the OTOC (for these system sizes $\Delta \tau \sim 1$).  

The right inset of  Fig~\ref{fig:4pt} shows
$J(\omega)$ vs $\omega$. In this case data for all system sizes collapse \emph{without} rescaling $\omega$ with $\ell$, as anticipated from Eq.~(\ref{Gform}) but in contrast to behaviour for $K_\ell(\omega)$. We expect in addition from Eq.~(\ref{Gform}) that $J(\omega) = F(\omega)$. In fact, the peak in $J(\omega)$ near $\omega=0$ is much more pronounced than is shown for $F(\omega)$ in Fig.~\ref{fig:2pt}. The discrepancy arises from different choices of boundary conditions: periodic for Fig.~\ref{fig:2pt}, open for Fig.~\ref{fig:4pt}. Viewed in the time domain, decay of $f(t)$ is slower for an operator at the end of an open chain than with periodic boundary conditions, because its spreading is hindered.
% $ \mathcal{I}(\omega_y) :=  \int d\omega_x  \, d \omega_z  \, G(\omega_x, \omega_y, \omega_z) $ against $\omega_x / \ell$ for $L=8,10,12$. This quantity corresponds to the Fourier transform of $\langle X(t/2)Y(t) X(t/2) Y(0)\rangle  \approx \langle Y(t)Y(0) \rangle$. 
The right inset of Fig.~\ref{fig:4pt} also shows that $F(\omega)$ for an operator at the end of an \emph{open} chain has a very similar form to $J(\omega)$.
%; we attribute the small differences to finite $\ell$.
%The peak is more prominent than the one in the top right inset of Fig.~\ref{fig:2pt} with periodic boundary condition.
%The peaks in the right insets are more prominent than the one in the top right inset of Fig.~\ref{fig:2pt} because in Fig.~\ref{fig:4pt}, the operator $\hat Y$ is located at the open edge of the system and its spreading is hindered. 
%In the bottom left inset, $\mathcal{I}'(\omega_z) :=  \int d \omega_x d \omega_y \;  H(\omega_x, \omega_y, \omega_z) $, where $H$ is defined by interchanging the first $X$ and $Y$ in $G$ in Eq.~\ref{G}. We see that $\mathcal{I}'(\omega_z)$ has a uniform profile. 

\begin{figure}[t!]
	\includegraphics[width=0.95\columnwidth]{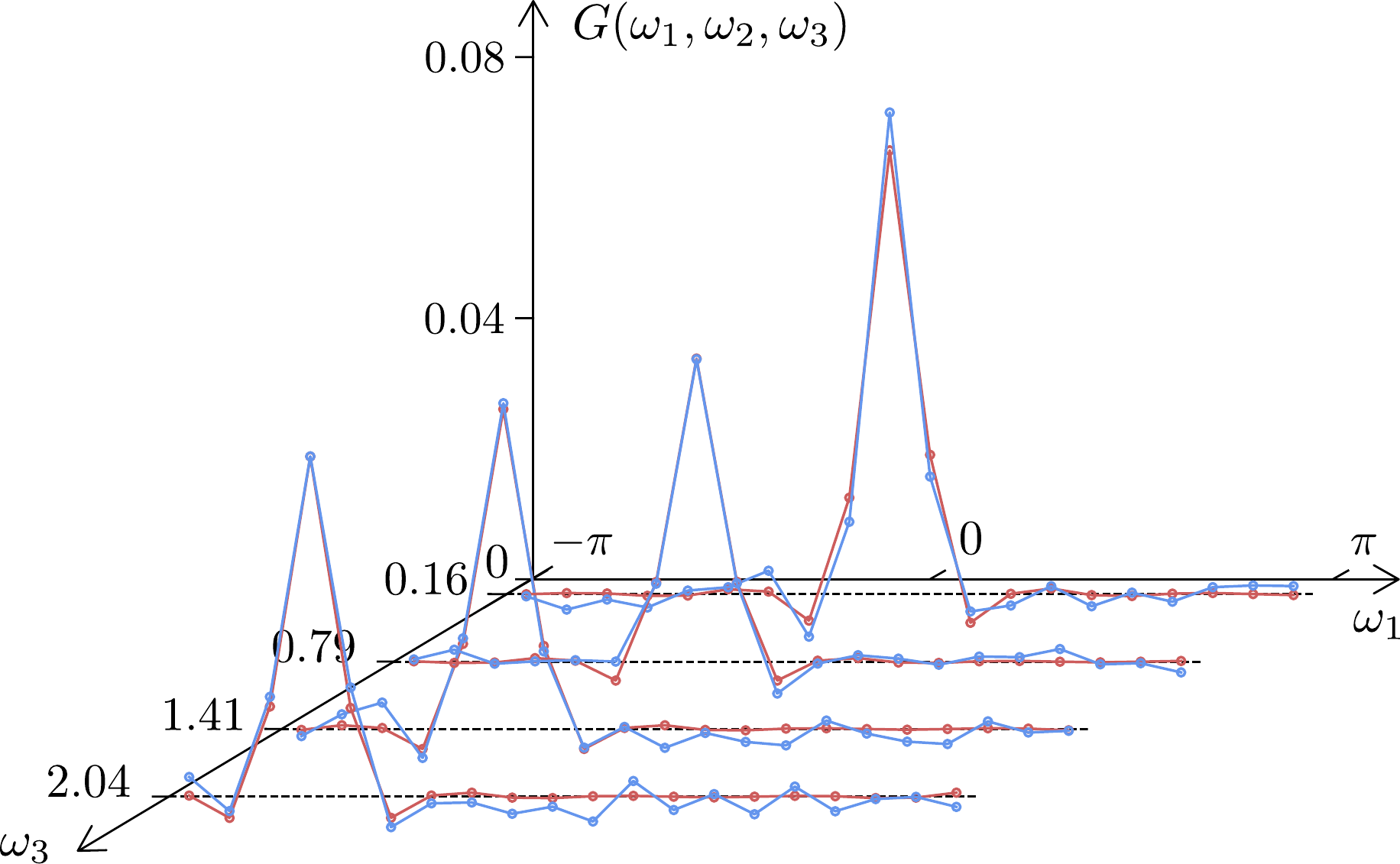} 
	\caption{
	To test Eq.~\ref{Gform}, we plot with offsets $G(\omega_1, \omega_2, \omega_3) $ (blue) and $F(\omega_2) F(\omega_3) K_\ell (\omega_1+ \omega_3)$ (red) vs $\omega_1$ for $L=12$ at $\omega_2 = -1.73$. % and $\omega_3 = 0.16, -0.49,-1.1, -1.73 $. 
			Plots for other values of $\omega_2$ and $\omega_3$ show similar agreement.
		%$ \mathcal{I}_{\omega_x, \omega_z}(\omega_y) := \int d \omega_x d \omega_z \;  G(\omega_x, \omega_y, \omega_z) $ vs. $\omega_y$. The small peak scales trivially with $\ell$ and originates from the auto-correlation of local operators at the edge of a system with OBC (see its Fourier transform in the bottom right inset). 
		%Bottom right: $F(\omega)$ vs $\omega$ for $\hat X$ acting on site at end of open chain for $L=??$. 
		%The peak is more prominent than the one in the top right inset of Fig.~\ref{fig:2pt} with PBC.
	} \label{fig:gvsffk}
\end{figure}

Finally, and crucially, we show support for our main result in  Fig~\ref{fig:gvsffk}. The product form given in Eq.~(\ref{Gform}) indeed provides a very accurate representation of $G(\omega_1, \omega_2, \omega_3)$.


\paragraph{Discussion.} 
It is natural to ask about the behaviour of other correlators. Within the Floquet model we have described, non-zero correlators must have even numbers of operators acting at each site, since odd powers vanish under the ensemble average. The only two-point correlator is hence $\langle \hat{X}(t)\hat{X}\rangle$. Besides the generalised OTOC, there is a second four-point correlator, with the form $\langle \hat{X}(t_3)\hat{X}(t_2) \hat{Y}(t_1) \hat{Y} \rangle$. This correlator has no long-time structure and therefore no small-frequency features. It is captured for large $\ell$ by the ETH, since it factorises in this limit as $\langle \hat{X}(t_3)\hat{X}(t_2) \rangle \langle\hat{Y}(t_1) \hat{Y} \rangle$. Hence $G(\omega_1, \omega_2, \omega_3) $ is unique in its sharp $\omega$-space structure. 
%This structure implies coherent contributions to  $X_{\alpha \beta} Y_{\beta \gamma} X_{\gamma \delta} Y_{\delta \alpha}$ if $(E_\alpha-E_\beta) \approx -(E_\gamma - E_\delta)$. 
%We also find [supplemental material] similar structure in generalisations of the OTOC at arbitrary higher (even) order.

We expect the phenomenology we have described to be very generic, as it arises from fundamental features of chaotic dynamics in spatially extended systems.
In particular, our conclusions will also hold in higher spatial dimensions. 
%Indeed, as clarified recenly in \cite{Nahum2017a}, operators are expected to spread ballistically though not necessarily isotropically, while the spatial dimensionality only affects the sub-linear broadening of the front in time. Therefore, the reasoning which led to Eqns.~(\ref{Gform}) and (\ref{k}) remains valid possibly including an angular dependence of the butterfly velocity.
In addition, although we have treated the simplified context of Floquet systems, we expect our conclusions to apply with some caveats in the presence of conserved quantities. To be specific, consider a system with a time-independent Hamiltonian $\hat H$ and energy as the only conserved density, 
%for which no other conserved quantities are present 
and examine matrix elements of operators between eigenstates of $\hat H$. In this case,
the appropriate thermal average at inverse temperature $\beta$ is $\aveeq{ \cdot }_{\beta} \equiv \mathcal{N}^{-1} \Tr[\,\cdot\, e^{-\beta \hat H}]$. A consequence, as for the standard ETH, is that all spectral correlators, including the functions $F$, $G$ and $K_\ell$, as well as $v_B$, acquire a smooth temperature dependence. Following \cite{rakovszky2017diffusive, khemani2018operator}, we expect that for operators $\hat X$ and $\hat Y$ that do not couple to the conserved charge, our conclusions will hold unchanged. On the other hand, if $\hat X$ or $\hat Y$ couple to the energy density, both the on-site relaxation in \eqref{autocorr} and the OTOC in \eqref{corr2} will present power-law tails.
%, consistently with the expectation that conserveddensities generically undergo diffusion. 
This slow dynamics may affect~\cite{Dymarsky} the separation of scales $\ell \gg v_B t_R$ that we exploited in deriving Eq.~\eqref{factorise} and we leave further analysis for future studies. 
 
%\paragraph{Conclusions. ---}

\paragraph{Acknowledgements.}
The work was supported in part by the European Union's
Horizon 2020 research and innovation programme under
the Marie Sklodowska-Curie Grant Agreement No.
794750 (A.D.L.), and in part by EPSRC Grant No. EP/N01930X/1 (A.C. and J.T.C.). We are grateful to S.~Gopalakrishnan, A.~Nahum and S.~Parameswaran for discussions.

\bibliography{FloquetChaos}

%%%%%%%%%%%%%%%%%%
%%%%%%%%%%%%%%%%%% Supplementary

\onecolumngrid
%\appendix
\newpage 

\appendix
\setcounter{equation}{0}
\setcounter{figure}{0}
\renewcommand{\thetable}{S\arabic{table}}
\renewcommand{\theequation}{S\arabic{equation}}
\renewcommand{\thefigure}{S\arabic{figure}}

\begin{center}
	{\Large Supplementary Material \\ 
		\titleinfo
	}
\end{center}
In this supplementary material we provide additional details about:
\begin{itemize}
	\item Computation of $G(\omega_1, \omega_2, \omega_3)$ under the standard hypothesis of ETH for the matrix elements of $\hat{X}$ and $\hat{Y}$.
	\item  The factorised form for the function $G(\omega_1, \omega_2, \omega_3)$.
\end{itemize}

\section{Computation of $G(\omega_1, \omega_2, \omega_3)$ under the standard hypothesis of ETH for the matrix elements of $\hat{X}$ and $\hat{Y}$}
	Here we analyse the  order of contributions in $G(\omega_1, \omega_2, \omega_3)$ that would be expected if the matrix elements of $\hat{X}$ and $\hat{Y}$ were described by the ETH in Eq.~\ref{ethmatrix}. Substituting Eq.~\eqref{ethmatrix} into \eqref{G} with $\overline{X}= \overline{Y}=0$, one can introduce the random variable
\begin{equation}\label{Gcaldef}
\mathcal{G}(\omega_1,\omega_2, \omega_3) = N^{-3} h(\omega_3) h(\omega_2) h(\omega_1) h(\omega_1 + \omega_2 + \omega_3) \sum_{\alpha\beta \gamma \delta} 
R_{\alpha \beta}^X R_{\beta \gamma}^Y R_{\gamma \delta}^X R_{\delta \alpha}^Y 
 \delta(\Delta_{\alpha\beta} - \omega_3)\delta(\Delta_{\beta\gamma} - \omega_2)\delta(\Delta_{\gamma\delta} - \omega_1)
\end{equation}
with $G(\omega_1, \omega_2, \omega_3) = \left[\mathcal{G}(\omega_1,\omega_2,\omega_3)\right]_{\rm{av}}$. %Here, we assume that $\delta$-functions smeared on the scale of level spacing. 
If the matrix elements of $\hat X$ and $\hat Y$ were independent gaussian random variables, one would have simply
\begin{equation}
 \left[ R_{\alpha \beta}^X R_{\gamma \delta}^X\right]_{\rm av} = \delta_{\alpha \delta} \delta_{\beta\gamma} \;, \qquad  \left[ R_{\beta \gamma}^Y R_{\delta \alpha}^Y\right]_{\rm av} = \delta_{\alpha \beta} \delta_{\delta\gamma}
\end{equation}
and hence the average
\begin{equation}
\label{Gave}
 G(\omega_1, \omega_2, \omega_3) \equiv \left[ \mathcal{G}(\omega_1, \omega_2, \omega_3) \right]_{\rm av} 
= N^{-2} [h(0)]^\john{4} \delta(\omega_3)\delta(\omega_2)\delta(\omega_1) \;.
\end{equation}
%As this term has a well-defined phase, we refer to it as the coherent contribution in the main text.
Under this hypothesis, $\mathcal{G}(\omega_1, \omega_2, \omega_3)$ is a random variable, whose behavior is characterised by its fluctuations. Indeed, by considering the $\mathcal{O}(N^4)$ terms in the sum where all indexes are different $\alpha \neq \beta \neq \gamma \neq \beta$ (other contributions are easily seen to be subleading in $N$), it follows from the central limit theorem that the variance of $\mathcal{G}(\omega_1, \omega_2, \omega_3)$ scales as
\begin{equation}
\label{Gvar}
 \mbox{Var}[\mathcal{G}(\omega_1,\omega_2, \omega_3)] = \mathcal{O}(N^{-2})\,.
\end{equation}

Eqs.~\eqref{Gave} and \eqref{Gvar} would imply that $\mathcal{G}(\omega_1, \omega_2, \omega_3)$
vanishes 
%with probability $1$ 
in the thermodynamic limit $N \to \infty$. Consider, however, the sum rule
\begin{align}\label{sum-rule}
\int_{[-\pi,\pi]^3}\!\!\!\!\!\!\!\! {\rm d} \omega_1 {\rm d}\omega_2 {\rm d}\omega_3 \,\mathcal{G}(\omega_1,\omega_2, \omega_3)& = N^{-1} \sum_{\alpha, \beta , \gamma, \delta} X_{\alpha \beta} Y_{\beta \gamma} X_{\gamma \delta} Y_{\delta \alpha} = 
\langle {\hat X}^2 {\hat Y}^2 \rangle = 1
\end{align}
This shows that the hypothesis of independence between matrix elements of $\hat X$ and $\hat Y$ cannot be correct. In particular, $X_{\alpha \beta} Y_{\beta \gamma} X_{\gamma \delta} Y_{\delta \alpha}  $ must have a coherent contribution of $\mathcal{O}(N^{-3})$ which, after averaging, 
% at sufficiently large $N$, 
dominates the incoherent fluctuation in \eqref{Gvar}. A quantitative determination of this contribution is one of the main results of our work, given in Eq.~\eqref{Gform} of the main text.

\section{Factorization of $G(\omega_1, \omega_2, \omega_3)$}
As discussed in the main text, the four-point correlator introduced in Eq.~\eqref{G} is related to $g(t_1, t_2, t_3)$
by the Fourier transform
\begin{equation}\label{eq:ft}
 G(\omega_1,\omega_2, \omega_3) =   \frac{1}{(2\pi)^3}\sum_{t_1, t_2, t_3} e^{\imath (\omega_1 t_1 + \omega_2 t_2 + \omega_3 t_3)}
 g(t_1, t_2, t_3) 
\end{equation}
where the sum is over the set of integers $(t_1, t_2, t_3) \in \mathbb{Z}^3$.  
Here, we show that $G(\omega_1, \omega_2, \omega_3)$ can be computed exactly in the limit $\ell \gg v_B t_R$ by replacing $g(t_1, t_2, t_3)$ with the factorised form in Eq.~\eqref{factorise}. 
We denote the corrections to \eqref{factorise} by
\begin{equation}
 r(t_1, t_2, t_3) = |g(t_1, t_2, t_3) - g_{\mbox{\tiny approx}}(t_1, t_2, t_3)| \;, \qquad 
 g_{\mbox{\tiny approx}}(t_1, t_2, t_3) = f(t_3 - t_1) f(t_2) k_{\ell}(\ell/v_B - |t_1|)
 \; .
\end{equation}
We set $\ell_{12} = \ell - |t_1 - t_2|/v_B$ and $\ell_{23} = \ell - |t_3 - t_2|/v_B$ and consider the possible regimes. 

To estimate $r(t_1, t_2, t_3)$, we discuss operator spreading using the notation of \cite{vonKeyserlingk2017}. For simplicity, we focus on the $q=2$ case, the generalisation to higher $q$ being straightforward. We denote Pauli operators acting on site $i$ as $\sigma_i^\mu$ with $\mu = 1,2,3$ and use $\sigma_i^0$ for the identity operator. 
We introduce Pauli strings as products of single-site Pauli operators $\hat{\mathcal{S}} = \prod_{i} \hat \sigma_i^{\mu_i}$. Then we can expand each Heisenberg operator as
	\begin{equation}
	\label{XYexp}
	 \hat X(t) = \sum_{\mathcal{S} } c_X^\mathcal{S}(t)  \hat{\mathcal{S}} \;, \qquad
	 \hat Y(t) = \sum_{\mathcal{S} } c_Y^\mathcal{S}(t)  \hat{\mathcal{S}} \;.
	\end{equation}
Note that Pauli strings form an orthonormal basis, since $\langle \mathcal{\hat{S}}^\dagger \mathcal{\hat{S}}' \rangle \equiv q^{-L}\Tr \mathcal{\hat{S}}^\dagger \mathcal{\hat{S}}'= \delta_{\mathcal{S} \mathcal{S}'}$. Because $\langle X^2 \rangle = \langle Y^2 \rangle = 1$, unitarity implies 
\begin{equation}
    \sum_{\mathcal{S}} |c^\mathcal{S}_X (t)|^2 = 1  \;, \qquad     \sum_{\mathcal{S}} |c^\mathcal{S}_Y (t)|^2 = 1  \;.
    \end{equation}
As the total weight of Pauli strings remain normalised, one can interpret their weight as a probability distribution. Two crucial features of a given string are the positions of its left and right ends: the left-most and right-most sites $i$ on which a Pauli operator $\sigma^\mu_i$ with $\mu=1, 2$ or $3$ appears, rather than the identity.  We use as guidance results on operator spreading known for random circuits that are stochastic in time. It has been shown \cite{vonKeyserlingk2017, Nahum2017a} at $t\gg t_R$ that  nearly all strings contributing to $\hat{X}(t)$ have their left end close to $x-v_Bt$ and their right end close to $x+v_B t$. The broadening of the fronts at $x\pm v_B t$ is diffusive in one dimension~\cite{vonKeyserlingk2017, Nahum2017a} and generically sub-ballistic. Between its left and right ends, a typical string has Pauli operators (as opposed to the identity) at a finite fraction of sites. We assume that similar behaviour also applies in spatially extended chaotic systems more generally. Using Eq.~\eqref{XYexp} we have
	\begin{equation} \label{eq:paulistringexp}
	g(t_1, t_2, t_3) =  \langle X(t_3) Y(t_2) X(t_1) Y \rangle  
	= 
	\sum_{\mathcal{S},\mathcal{S}', \mathcal{S}''} 
			c_X^{\mathcal{S}''}(t_3) c_Y^{\mathcal{S}'}(t_2) c_X^\mathcal{S}(t_1)
	\langle \mathcal{S}'' \mathcal{S}'  \mathcal{S} Y \rangle   \; .
	\end{equation}
The average $\langle \mathcal{S}'' \mathcal{S}'  \mathcal{S} Y\rangle$ is unity if the product of operators $\sigma_i^\mu$ from the four strings $\mathcal{S}''$, $\mathcal{S}'$, $\mathcal{S}$ and $Y$ gives the identity at every site $i$; otherwise it is zero. 

\begin{figure}[h!]
		\centering
			\includegraphics[angle=0, width=0.65\linewidth]{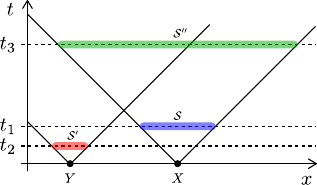}
			\caption{Sketch of the spreading of operators $X(t_1)$ (blue), $Y(t_2)$ (red) and $X(t_3)$ (green). Most of the weight in the expansion \eqref{XYexp} is concentrated in strings which have a support growing ballistically with time at the butterfly velocity $v_B$. }\label{operatorsLC}
	\end{figure}

The typical form of these four strings is shown schematically in Fig.~\ref{operatorsLC}. The dominant contributions to $g(t_1, t_2, t_3)$ arise from sets of strings in which all three of the following conditions are met: (i) the left and right ends of $\mathcal{S}''$ coicide with those of $\mathcal{S}$; (ii) $\mathcal{S}'$ coincides with $Y$; and (iii) the site $y$ is not situated between the left and right ends of $\mathcal{S}$. 
    
To formalise this approach, we introduce the total weight of Pauli strings laying to the left of site $i$ as
    \begin{equation}
    \label{Rdef}
     R_X(i, t) \equiv \sum_{\substack{\mathcal{S}\, | \, \operatorname{RHS}(\mathcal{S})<i}} |c^\mathcal{S}_X (t)|^2 
    \end{equation}
    where $\operatorname{RHS}(\mathcal{S})$ labels the right end of the Pauli string $\hat{\mathcal{S}}$. The operator $\hat X$, initially localised at the site $x$, spreads ballistically at velocity $v_B$ and one expects at large times~\cite{vonKeyserlingk2017, Nahum2017a}
    \begin{equation}
    \label{Rspreading}
        R_X(i,t) \sim 1 - C_{\ell = |i - x|}(t) = \begin{cases}
                                                   0 & i \ll x + v_B t\\
                                                   1 & i \gg x + v_B t
                                                  \end{cases}
\;\;.
    \end{equation}
with corrections that decrease with the distance $|i - x - v_B t|$ from the front. 

We discuss \eqref{factorise} in separate steps, according to the values of $\ell_{12}$ and $\ell_{23}$. Consider first $|\ell_{12}|$ and $|\ell_{23}|$ small, and assume for definiteness diffusive front broadening with diffusion constant $D$. Then we expect deviations from the factorised form \eqref{factorise} for $|\ell_{12}| \lesssim  O(\sqrt{D \ell/v_B})$  (or  $|\ell_{23}| \lesssim  O(\sqrt{D\ell/v_B})$), because in this case the ends of strings $\mathcal{S}'$ and $\mathcal{S}$ (or $\mathcal{S}'$ and $\mathcal{S}''$) are close. However, because of relaxation arising from scrambling, $r(t_1, t_2, t_3)$ can be $O(1)$ only if $|t_1 - t_3| \lesssim O(t_R)$ and $t_2 \lesssim O(t_R)$. Upon integration, this gives a contribution from $r(t_1,t_2,t_3)$ of size $O(t_R^2  \sqrt{\ell })$ to $ G(\omega_1,\omega_2, \omega_3) $, which for large $\ell$ is parametrically smaller than the contribution described in the main text.

Next we examine the regimes where $|\ell_{12}|$ and $|\ell_{23}|$ are both large and 
consider the possible cases. When  $\ell_{12}$ and $\ell_{23}$ are both large and positive, the operators $X(t_1)$ and $Y(t_2)$ in \eqref{defsmallg} commute and the expectation value factorises. We have $g(t_1, t_2, t_3 ) = g_{\mbox{\tiny approx}}(t_1,t_2,t_3)= f(t_1- t_3) f(t_2)$ and so $r(t_1,t_2,t_3) = 0$ identically.

In the other cases, it is easy to convince oneself that for a generic system both $g_{\mbox{\tiny approx}}(t_1, t_2, t_3)$ and $g(t_1, t_2, t_3)$ are simultaneously small. To be specific, consider the case when $\ell_{12}>0$ and $\ell_{23}<0$, both being large in absolute value. 
Clearly, $g_{\mbox{\tiny approx}}(t_1, t_2, t_3)$ vanishes for $|t_1 - t_3| \gg t_R$ because of the factor $f(t_3 - t_1)$. To estimate $g(t_1, t_2, t_3)$ we use Eq.~(\ref{eq:paulistringexp}), taking terms in the sum for which $\langle \mathcal{S}'' \mathcal{S}'  \mathcal{S} Y\rangle=1$. Because of $\eqref{Rspreading}$, most of the weight of the operator $X(t_1)$ is concentrated in strings with support spread throughout the interval $[x - v_B t_1, x + v_B t_1]$. However, in the regime under consideration, $t_3 \gg t_1$, and so non-vanishing contributions to \eqref{eq:paulistringexp} come from atypical strings $\mathcal{S}''$ which are anomalously short (see Fig.~\ref{operatorsLC}). More explicitly, we restrict the sum over the strings $\mathcal{S}''$ which overlap with $\mathcal{S}$ and $\mathcal{S'}$
	\begin{multline}
	|g(t_1, t_2, t_3)|	= 
	\left|\sum_{\mathcal{S},\mathcal{S}'} \sum_{\substack{\mathcal{S}'' \\ \mbox{\tiny overlap}}} 
			c_X^{\mathcal{S}''}(t_3) c_Y^{\mathcal{S}'}(t_2) c_X^\mathcal{S}(t_1)\right| \leq \left(\sum_{\substack{\mathcal{S''} \\  \operatorname{RHS}(\mathcal{S''})<x + v_B t_1}} |c^\mathcal{S''}_X (t_3)|^2\right)^{1/2} \left(\sum_{\mathcal{S}, \mathcal{S}'} |c^\mathcal{S}_X (t_1)|^2 |c^\mathcal{S'}_{Y} (t_2)|^2\right)^{1/2} = \\= R_X(x + v_B t_1, t_3)  .	 
	\end{multline}
where we used Cauchy-Schwartz inequality and the definition \eqref{Rdef}. So, because of \eqref{Rspreading}, $g(t_1, t_2, t_3)$ is small in this regime. The other regimes can be addressed in a similar way.

\end{document}